# Oxidation-induced ultrafast spin-to-orbital conversion at heavy-metal interfaces

Xiaoxue Zeng[1], Tianyi Zhang[2], Yaokai Niu[1], Qiye Wen[1], Zhiyong Zhong[1], Zhi-Min Liao[3], Peng Yan[1], Xiufeng Han[2,*] and Lichuan Jin[1,†]

[1] State Key Laboratory of Electronic Thin Films and Integrated Devices, University of Electronic Science and Technology of China, Chengdu, Sichuan 610054, China

[2] Beijing National Laboratory for Condensed Matter Physics, Institute of Physics, Chinese Academy of Sciences, Beijing 100190, China

[3] State Key Laboratory for Mesoscopic Physics and Frontiers Science Center for Nano-optoelectronics, School of Physics, Peking University, Beijing 100871, China

Oxidation engineering provides a route to control orbital degrees of freedom, yet its role in spin-to-orbital conversion remains largely unexplored. Here, we report an efficient spin-to-orbital conversion mechanism driven by interfacial oxidation at heavy-metal interfaces. In W/Co/$SiO_2$ heterostructures, terahertz emission exhibits a time delay that scales linearly with the W thickness, identifying orbital-current transport as the dominant origin. The emission amplitude is approximately three times larger than that of Co/Pt bilayers, indicating highly efficient conversion from spin to orbital angular momentum. Systematic variation of Co thickness, stoichiometry, and interface configuration reveals that the effect originates from oxidation of the W layer at the W/Co interface, which modulates the interfacial orbital texture. We further show that this mechanism is generic across different heavy metals and scales with their spin-orbit coupling strength. These results establish oxidation as an effective handle to engineer spin-to-orbital conversion and provide a general route toward orbitronic terahertz emitters.

Orbitronics, which exploits the orbital degrees of freedom of electrons as information carriers, represents a natural extension of spintronics and offers a promising physical platform for next-generation low-power and high-performance devices [1-4]. In conventional spintronic terahertz (THz) emitters, femtosecond laser excitation of a ferromagnetic (FM) layer generates a transient spin current that is injected into an adjacent nonmagnetic (NM) layer [5,6] and converted into an ultrafast charge current via mechanisms such as the inverse spin Hall effect [7] or the inverse Rashba-Edelstein effect [8-11], resulting in broadband THz radiation [12]. While this spin-based approach has enabled efficient THz sources, its performance is fundamentally limited by the efficiency of spin-to-charge conversion. Recent studies have revealed that orbital angular momentum plays an essential role in ultrafast transport and angular-momentum conversion processes, suggesting that orbital currents can serve as an intermediate or even dominant channel beyond purely spin-based schemes[13-15]. This emerging understanding has stimulated growing interest in orbitronic-based THz emitters, with proposed strategies including light/heavy metal heterostructures [16,17], rare-earth-element [18] or FM layer [19] engineering, and alloy systems such as CoPt [20,21]. Despite these advances, the generation, conversion, and transport mechanisms of orbital angular momentum under femtosecond excitation remain far from fully understood, leaving substantial room for improving both emission efficiency and device design.

Controlled oxidation has recently emerged as an effective strategy for manipulating interfacial symmetry, orbital hybridization, and electronic structure in spintronics and orbitronics[22,23]. In spintronic and orbitronic devices, interfacial oxidation has been widely demonstrated to significantly enhance charge-to-orbit conversion efficiency. This improvement is generally attributed to oxidation-induced interfacial reconstruction and the redistribution of orbital-related electronic states [24-29]. Oxidation engineering also provides a promising route for improving THz emission efficiency. Recent studies have shown that introducing a light-metal oxide layer into multilayer heterostructures, thereby supplementing the orbital current contributions, can effectively enhance the THz emission amplitude [16,19,20]. Therefore, a clear understanding of how controlled interfacial oxidation under femtosecond excitation modulates orbital states and spin-orbit angular momentum conversion is essential for advancing orbitronic terahertz emitters. However, despite the potential significance of oxidation engineering in orbitronic THz systems, existing research has primarily focused on orbital current generation, while its influence on spin-to-orbital conversion remains largely unexplored.

In this work, we report an exceptionally efficient spin-to-orbital conversion enabled by controllable heavy-metal interfacial oxidation in W/Co/$SiO_2$ heterostructures under femtosecond excitation, manifested as a pronounced enhancement of terahertz emission amplitude. By systematically tuning the layer thickness, material composition, and interfacial configuration, we show that this enhancement emerges exclusively in sufficiently thin Co layers and is dominated by orbital angular momentum transport and conversion. We further demonstrate that the terahertz amplitude enhancement is universal across heavy metals with strong spin-orbit coupling, while both the signal polarity and conversion efficiency are dictated by the intrinsic L-S correlation of the heavy-metal layer; a mildly oxidized interface acts as the decisive trigger for activating this ultrafast conversion channel. These findings establish interfacial oxidation as a powerful handle for engineering nonequilibrium angular momentum flow and provide a new design paradigm for orbitronics-based terahertz emitters, opening a viable pathway toward high-efficiency ultrafast terahertz sources rooted in orbital physics.

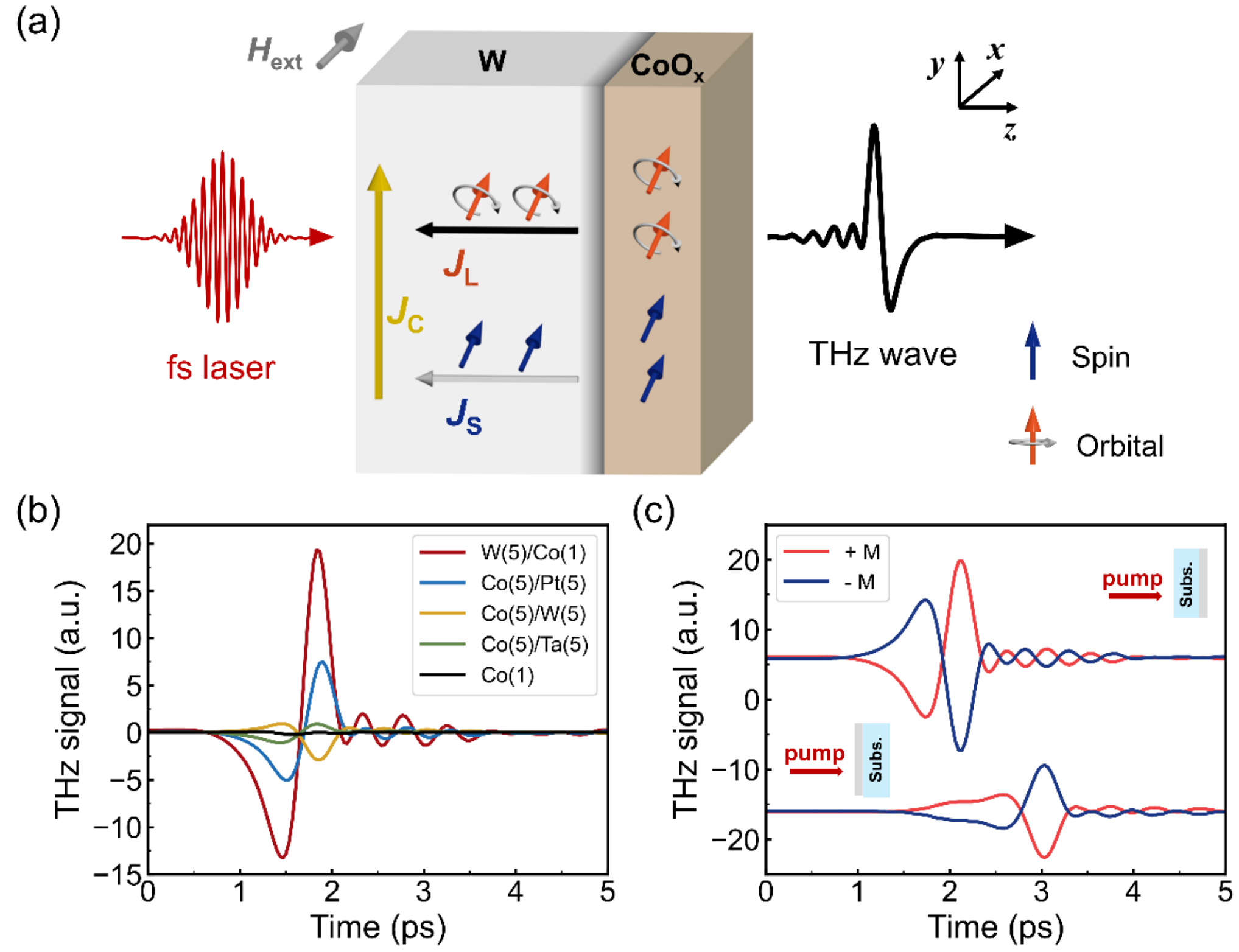

FIG. 1. (a) Schematic illustration of the THz emission from W/Co/$SiO_2$ heterostructure. An in-plane external magnetic field of 700 Oe is applied during the measurements. (b) THz emission signals from W(5)/Co(1)/$SiO_2$, Co(5)/NM(5)/$SiO_2$ (with NM=Pt, W, Ta), and Co(1)/$SiO_2$ heterostructures measured under identical experimental conditions. The numbers in parentheses indicate the layer thicknesses in nanometers, and for clarity, "$SiO_2$(3)" is omitted from the labels of all sample structures. (c) THz emission signals from W(3)/Co(1)/$SiO_2$(3) heterostructure measured under reversed in-plane magnetization directions and opposite excitation geometries.

The investigated NM/Co and Co/NM heterostructures were deposited on $SiO_2$ substrates (thickness 0.5 mm) via dc magnetron sputtering, where NM=W, Pt, Ta, Ti, or Cu. Unless otherwise specified, all samples were capped with a 3 nm thick $SiO_2$ layer. We measured the terahertz signals of different samples using the THz emission spectroscopy [30]. The experimental configuration is schematically illustrated in Fig. 1(a). Femtosecond laser pulses were incident from the $SiO_2$ substrate side and excite the ferromagnetic Co layer. Photoexcitation generates ultrafast angular momentum pulses in Co, which propagate into the adjacent nonmagnetic layer and are converted into transient charge currents via spin- or orbital-to-charge conversion processes, leading to the emission of THz radiation. Fig. 1(b) compares the THz emission signals from representative heterostructures. Co(5)/NM(5)/$SiO_2$ heterostructures (NM = Pt, W, or Ta; thicknesses in nanometers) exhibit THz emission amplitudes comparable to those of conventional spintronic emitters [31,32]. However, a pronounced enhancement of the THz emission is observed when the stacking order is reversed to form a W(5)/Co(1)/$SiO_2$ heterostructure. In particular, the THz signal from W(5)/Co(1)/$SiO_2$ reaches a peak amplitude approximately three times larger than that of the Co(5)/Pt(5)/$SiO_2$ bilayer measured under identical conditions.

To reveal the physical origin of the THz emission, we examine its symmetry properties. As shown in Fig. 1(c), reversing the in-plane magnetization while keeping the sample orientation fixed leads to a polarity reversal of the THz signal, with the amplitude remaining nearly unchanged. An analogous polarity reversal is observed when the sample orientation is reversed under a fixed magnetic field. These symmetry relations indicate that the photoexcited angular momentum originates from the Co layer and that the THz emission arises from structural symmetry breaking, consistent with an electric-dipole origin of the THz emitter [33,34]. Such behavior is characteristic of spin- or orbital-mediated THz emission. A modification of the waveform upon sample reversal is attributed to asymmetric optical absorption in the FM/NM bilayer and to THz propagation and dispersion effects in the $SiO_2$ substrate [33].

Despite exhibiting the same symmetry rules as conventional spintronic THz emitters, the magnitude of the THz emission from W(5)/Co(1)/$SiO_2$ is anomalous. Within existing frameworks, reducing the thickness of the ferromagnetic layer is expected to decrease the total photoexcited angular momentum [31]. Moreover, excitation from the W side introduces substantial optical absorption in the W layer, further attenuating the laser energy reaching the Co layer [6]. Consistently, the Co(1)/$SiO_2$ reference sample shows nearly negligible THz emission (see Fig. 1(b)).

These observations indicate that the exceptionally strong THz signal observed in W(5)/Co(1)/$SiO_2$ cannot be explained by established THz emission mechanisms, pointing to the emergence of a distinct and unconventional angular-momentum conversion process.

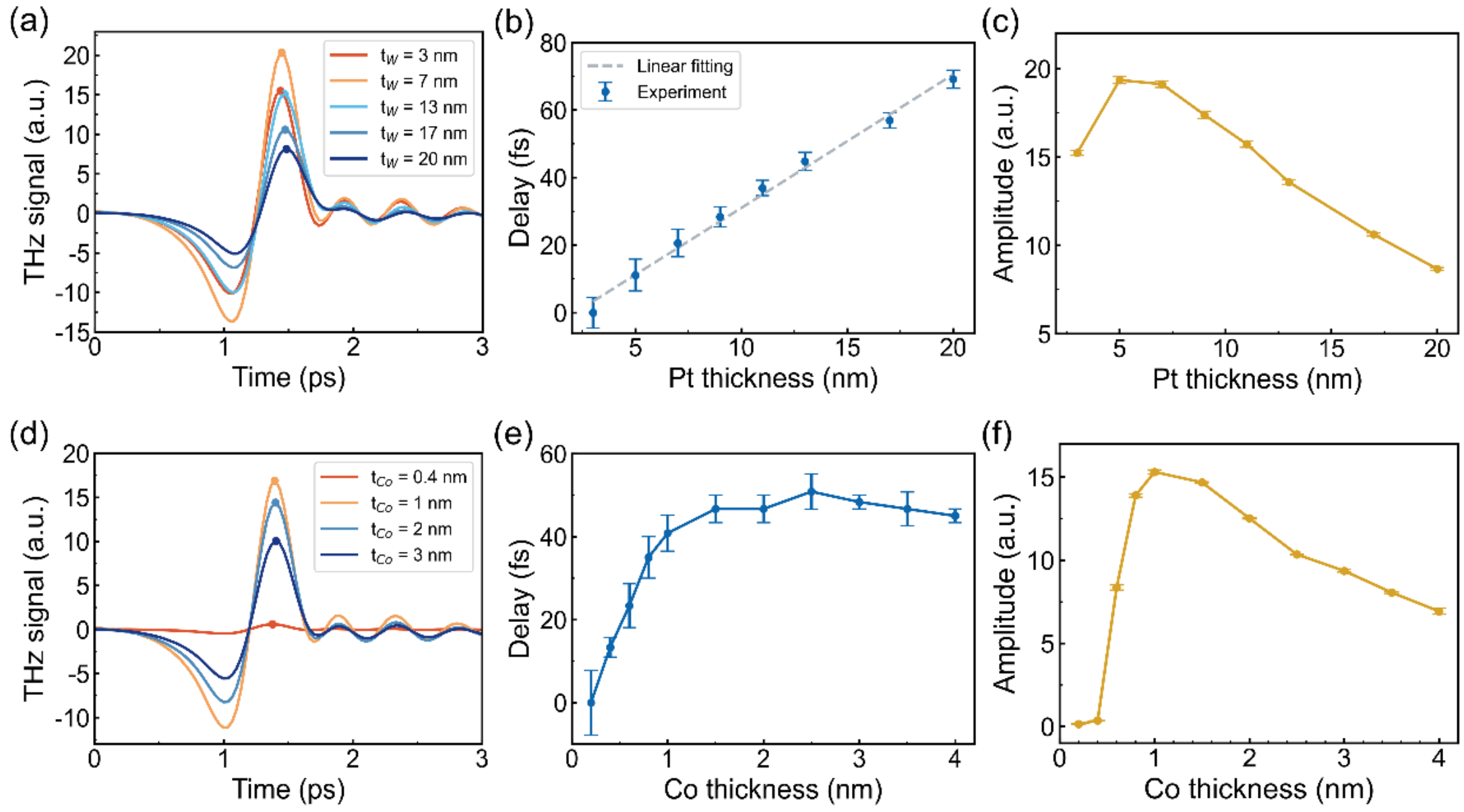

FIG. 2. (a) THz emission signal from W($t_W$)/Co(1)/$SiO_2$ heterostructures with varying W thickness $t_W$, while all other experimental conditions are kept unchanged. Peaks of the THz signals are indicated with circles. (b) Extracted THz peak time delay relative to the sample with $t_W$ = 3 nm, together with a linear fit. (c) Peak amplitude of THz signal as a function of $t_W$. (d) THz emission signals from W(3)/Co($t_{Co}$)/$SiO_2$ heterostructures with $t_{Co}$ varying from 0.2 to 4 nm. (e, f) Extracted THz peak time delay in (e) and amplitude in (f), respectively, as a function of $t_{Co}$.

Symmetry analysis indicates that the THz emission in the W/Co/$SiO_2$ heterostructure originates from the W layer. To investigate the angular momentum transport in W, we measured the THz emission as a function of the W thickness ($t_W$), as shown in Fig. 2(a). The peak position of each THz waveform was extracted using a Python-based analysis, where the horizontal and vertical axes correspond to the time delay and signal amplitude, respectively. Figure 2(b) presents the extracted peak delay as a function of $t_W$. Within the measured thickness range up to 20 nm, the delay increases linearly with $t_W$. This linear dependence indicates a long-range, ballistic transport of angular momentum in W, implying that the transported angular momentum is predominantly orbital in nature and that the orbital-to-charge conversion occurs at the W/oxide interface [21]. Furthermore, the time delay of the transmitted THz signal exhibits a nonlinear dependence on pump fluence, initially increasing and subsequently decreasing. This behavior is consistent with previous reports showing that orbital-current transport can be modulated by the coupling between orbital angular momentum and the crystal field, which is also a characteristic feature of ultrafast orbital angular momentum transport [30]. For comparison, we measured the THz emission from W($t_W$)/Co(5)/$SiO_2$ samples. With a Co thickness of 5 nm, the W/Co interface is effectively free of oxidation. The THz emission delay remains nearly constant with increasing W thickness ($t_W$), consistent with the short spin diffusion length in W [30].

This finding challenges the conventional understanding that the angular momentum pulse generated by femtosecond laser excitation in ferromagnetic Co is dominated by spin angular momentum, even in the presence of finite spin-orbit coupling [13]. As shown in Fig. 1(b), the THz waveforms from Co(5)/Pt(5) and Co(5)/W(5) heterostructures exhibit opposite polarities, consistent with spin-dominated transport. This observation indicates an exceptionally high spin-to-orbital conversion efficiency in the W/Co/$SiO_2$ heterostructure. In contrast, the Co(5)/Ta(5) heterostructure displays an anomalous polarity, which may originate from the intrinsic spin-orbit coupling of Co combined with the substantially larger orbital Hall conductance of Ta compared to its spin Hall conductance [4]. In addition, the dependence of the THz amplitude on $t_W$ follows a trend similar to previous reports: the signal reaches a maximum at approximately $t_W$=5 nm and gradually decreases for larger thicknesses, as shown in Fig. 2(c). We analyzed the orbital transport process in the W layer using theoretical and numerical methods (Ref. [30]).

Figure 2(d) shows the dependence of the THz emission signal on the Co layer thickness $t_{\mathrm{Co}}$=2-4 nm. By extracting the temporal position of the THz peak, we obtain the thickness dependence of the emission delay, as shown in Fig. 2(e). The delay increases with $t_{Co}$ for ultrathin Co layers ($t_{Co}\leq1$ nm) and subsequently saturates for thicker films. This behavior may arise from incomplete intermixing or partial segregation of interfacial elements when the Co layer is extremely thin. Once the Co thickness exceeds approximately 1 nm, the interfacial contribution to the time delay saturates, thereby preventing any further increase in the THz emission delay. The THz emission amplitude exhibits a thickness dependence that is markedly different from trends reported in previous studies (Fig. 2(f)) [31]. For $t_{Co}>0.4$ nm, the amplitude initially increases rapidly with increasing Co thickness, reaches a maximum at approximately $t_{Co}=1$ nm, and then gradually decreases, approaching the characteristic amplitude observed in conventional spintronic THz emitters. For comparison, we also measured the THz emission from the Co(1)/W(3)/$SiO_2$ heterostructure under identical experimental conditions [30]. The measured THz amplitude, drastically lower than that of the W/Co/$SiO_2$ structure and reverting to the level of a conventional spintronic emitter, confirms that the abnormal enhancement did not result from the formation of CoW alloy via interfacial mixing. These findings demonstrate that the efficient spin-to-orbital conversion is highly dependent on the interfacial structure.

To elucidate the origin of the spin-to-orbital conversion mechanism, we first replaced the NM layer with Pt, Ta, Ti, and Cu. As shown in Fig. 3(a), a pronounced enhancement of the THz emission is observed only for HM with strong spin-orbit coupling, such as Pt and Ta, whereas the signal remains weak for light metals (Ti and Cu). In addition, the THz signal polarity for W is the same as that for Ta, and opposite to that of Pt, consistent with their spin-orbit-related response [35,36]. These observations indicate that the efficient spin-to-orbital conversion is governed by the spin-orbit coupling of the NM layer rather than by the Co layer itself. To examine the role of interfacial effects, a 1 nm Cu spacer was inserted at the W/Co and Co/$SiO_2$ interfaces. As shown in Fig. 3(b). The THz amplitude in the Cu-inserted samples does not exhibit a pronounced reduction, suggesting that neither the W/Co nor the Co/$SiO_2$ interface plays a dominant role in determining the THz emission strength. This result implies that the spin-to-orbit conversion is not confined to a sharply defined HM/FM interface, but is instead associated with modifications at the heavy-metal surface.

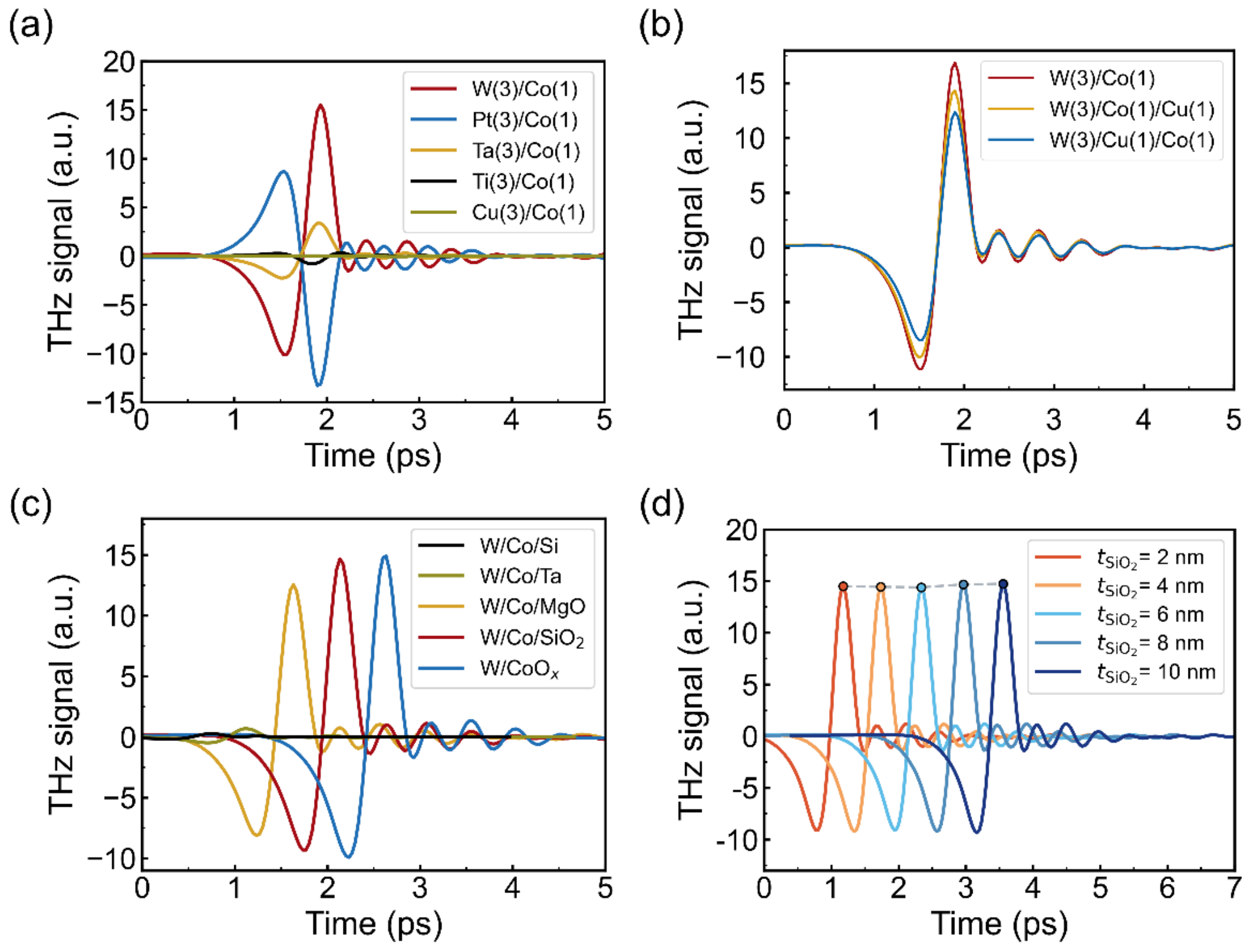

FIG. 3. (a) THz emission signal from NM/Co/$SiO_2$ heterostructure, where NM=W, Pt, Ta, Ti, or Cu. The layer thicknesses, given in nanometers, are indicated in parentheses. (b) THz emission signal from W/Co, W/Co/Cu, and W/Cu/Co heterostructures. (c) THz emission signal from W(3)/Co(1)/M(3) heterostructures, where M = Si, Ta, MgO, $SiO_2$, or no capping layer. (d) THz emission signals from W(3)/Co(1)/$SiO_2$($t_{SiO_2}$) and the extracted their peak amplitude as a function $t_{SiO_2}$.

Combined with the previous observation that the enhancement disappears upon reversing the stacking order and requires an ultrathin Co layer, these results point to a mechanism that is highly sensitive to the chemical environment of the heavy-metal surface. A plausible origin is interfacial oxidation. To test this hypothesis, we replaced the $SiO_2$ capping layer with Si, Ta or MgO, and examined samples without any capping layer for comparison. Si or Ta, which have negligible spin Hall conductivity and effectively suppress oxidation, are expected to isolate oxide effects. The results are summarized in Fig. 3(c) (Ref. [4]). While samples capped with MgO or $SiO_2$ exhibit THz amplitudes comparable to those of naturally oxidized samples, the THz signal is strongly suppressed when Si or Ta is used as the capping layer. This behavior demonstrates that the enhancement of THz emission in the W/Co/$SiO_2$ heterostructure is enabled by oxidation at the heavy-metal interface. Finally, we systematically varied the thickness of the $SiO_2$ capping layer, as shown in Fig. 3(d). The THz amplitude remains essentially unchanged with increasing $SiO_2$ thickness, indicating that the oxygen involved in the enhancement does not originate from ambient exposure. Instead, it is attributed to an elemental intermixing between the $SiO_2$ layer and the underlying structure. This result further suggests that the oxidation-induced enhancement is localized and stable under the present experimental conditions.

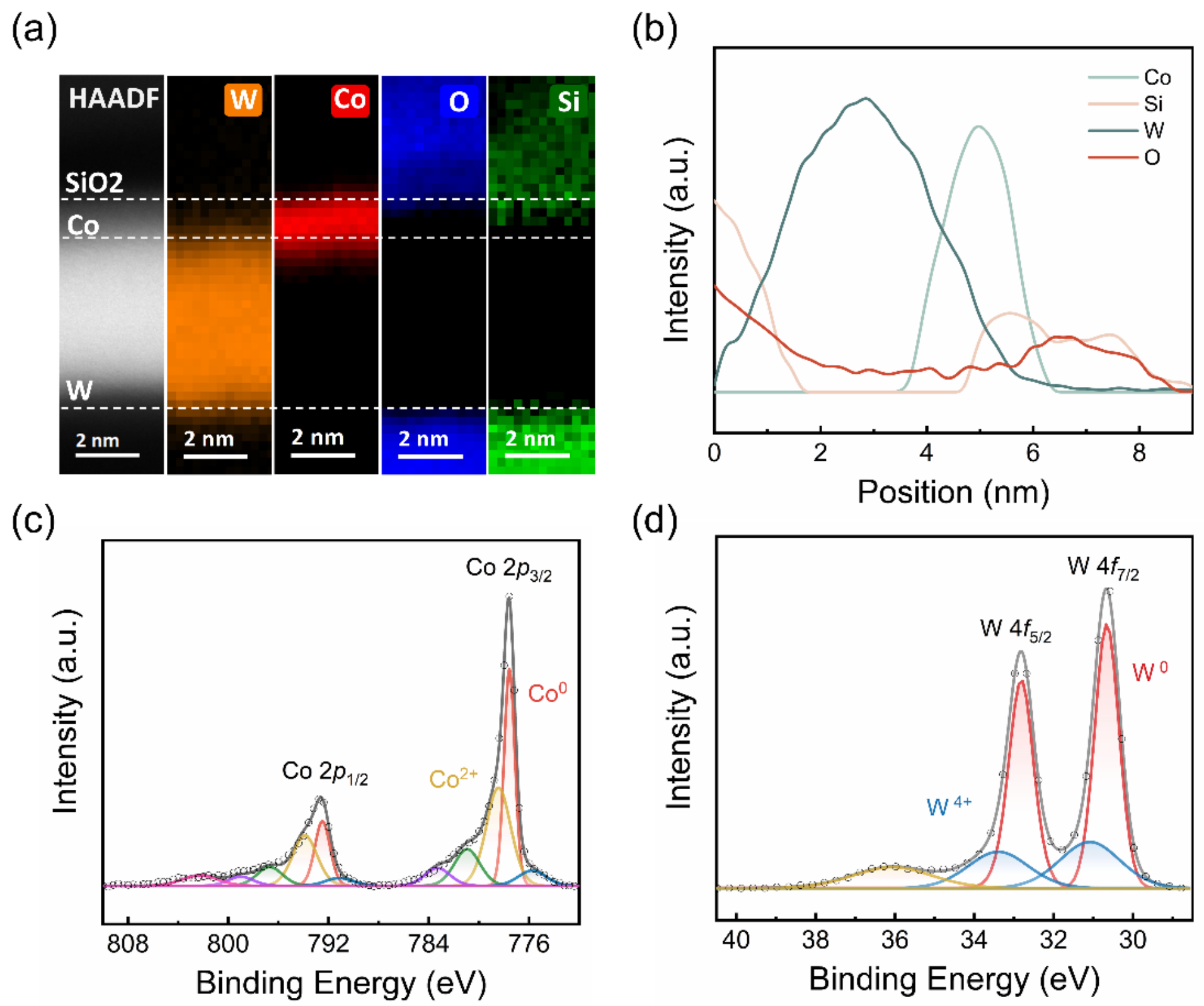

FIG. 4. (a) Cross-sectional HAADF-STEM image of the W(5)/Co(1)/$SiO_2$ heterostructure with the corresponding EELS elemental mappings. (b) EELS line scanning of Co, Si, W, and O along the thickness direction of the sample. (c-d) High-

resolution XPS of (c) Co and (d) W for the W(5)/Co(1)/$SiO_2$ heterostructure.

To verify the oxidation state of the thin-film structure, we performed detailed structural and chemical characterizations of the W(5)/Co(1)/$SiO_2$ heterostructure. Figure 4(a) shows a cross-sectional high-angle annular dark-field scanning transmission electron microscope (HAADF-STEM) image together with the corresponding electron energy loss spectrum (EELS) elemental mappings of W, Co, O, and Si. A well-defined multilayer structure is observed, while noticeable elemental intermixing occurs at the interfaces. Owing to the ultrathin Co layer, both W and $SiO_2$ partially overlap within the Co region. This interfacial intermixing is further confirmed by the EELS line profiles along the thickness direction shown in Fig. 4(b), which reveal a contact between the W and $SiO_2$ layers within the Co layer. As a result, oxygen can directly reach the surface of the W layer. This behavior is in clear contrast to the Co(5)/W(5)/$SiO_2$ structure, where the layers are well separated [30], indicating that such intermixing is suppressed when the ferromagnetic layer is sufficiently thick.

The chemical states of Co and W were further examined by X-ray photoelectron spectroscopy (XPS). As shown in Fig. 4(c) and Fig. 4(d), the Co 2p spectrum exhibits pronounced peak broadening, indicating significant oxidation of the Co layer [37,38]. In contrast, the W 4f spectrum is dominated by the metallic W component, with no detectable $WO_3$ peak [39-41]. Only a minor suboxide contribution is observed, consistent with oxidation localized near the interface. Note that the unmarked small peaks are satellite peaks. In addition, XPS characterization of the W(5)/Cu(1)/Co(1)/$SiO_2$ heterostructure was carried out [30]. The fitting results reveal partial oxidation of both the Co and Cu layers, while the oxidation state of the W layer remains comparable to that of samples without Cu insertion. This indicates that a 1 nm Cu spacer cannot fully suppress oxidation. Consistently, this explains why the THz emission amplitude does not exhibit a pronounced reduction upon Cu insertion, in agreement with the results shown in Fig. 3(b).

We discuss the spin-to-orbital conversion mechanism. Oxidation of the heavy metal layer at the W/Co interface modifies the interfacial orbital texture via oxygen-induced hybridization, enabling efficient interfacial spin-to-orbital conversion. The resulting orbital-polarized current is governed by the intrinsic L-S correlation of the heavy metal: the orbital polarization follows ($\eta_{L\text{-}S}>0$, e.g., Pt) or opposes ($\eta_{L\text{-}S}<0$, e.g., W and Ta) the spin polarization. The conversion efficiency is strongly dependent on interfacial oxidation and is suppressed when oxidation is inhibited or the stacking order is reversed. In addition, we also verified the performance of W/Co/$SiO_2$ heterojunction as an integrable terahertz source[30].

In summary, we demonstrate efficient spin-to-orbital conversion induced by interfacial oxidation in W/Co/$SiO_2$ heterostructures. This effect emerges for ultrathin Co layers and is accompanied by a W-thickness-dependent time delay, indicating orbital-dominated transport. Measurements show that the effect is prevalent in heavy metals, with the terahertz emission polarity and amplitude governed by their L-S correlation, whereas light metals exhibit no comparable response. Interface engineering further reveals that the conversion efficiency is controlled by oxidation of the heavy-metal interface, with the Co thickness directly influencing the degree of oxidation at the W surface. In addition, the HM/oxide interface provides a strong orbital-to-charge response, leading to terahertz amplitudes up to three times that of conventional Co/Pt bilayers. These results highlight interfacial oxidation as an effective means to modulate spin and orbital degrees of freedom, providing a viable route toward high-efficiency terahertz emitters.

## ACKNOWLEDGEMENTS

This work was financially supported by the National Key Research and Development Plan under Grant No. 2022YFA1402802; the National Natural Science Foundation of China under Grant No. 62171096; and the Natural Science Foundation of Sichuan Province under Grant No. 2026NSFSC0384.

## DATA AVAILABILITY

The data that support the findings of this article are not publicly available upon publication because it is not technically feasible and/or the cost of preparing, depositing, and hosting the data would be prohibitive within the terms of this research project. The data are available from the authors upon reasonable request.

†Contact author: lichuanj@uestc.edu.cn (L. Jin)
*Contact author: xfhan@iphy.ac.cn (X. Han)

[1] S. Zhang ,and Z. Yang, Intrinsic spin and orbital angular momentum Hall effect. Phys. Rev. Lett. **94**, 066602 (2005).

[2] D. Go, D. Jo, C. Kim, and H.-W. Lee, Intrinsic Spin and Orbital Hall Effects from Orbital Texture. Phys. Rev. Lett. **121**, 086602 (2018).

[3] D. Jo, D. Go, and H.-W. Lee, Gigantic intrinsic orbital Hall effects in weakly spin-orbit coupled metals. Phys. Rev. B **98**, 214405 (2018).

[4] D. Lee, D. Go, H.-J. Park, W. Jeong, H.-W. Ko, D. Yun, D. Jo, S. Lee, G. Go, J. H. Oh *et al.*, Orbital torque in magnetic bilayers. Nat. Commun. **12**, 6710 (2021).

[5] M. Battiato, K. Carva, and P. M. Oppeneer, Superdiffusive Spin Transport as a Mechanism of Ultrafast Demagnetization. Phys. Rev. Lett. **105**, 027203 (2010).

[6] A. Eschenlohr, M. Battiato, P. Maldonado, N. Pontius, T. Kachel, K. Holldack, R. Mitzner, A. Föhlisch, P. M. Oppeneer, and C. Stamm, Ultrafast spin transport as key to femtosecond demagnetization. Nat. Mater. **12**, 332 (2013).

[7] T. Kampfrath, M. Battiato, P. Maldonado, G. Eilers, J. Nötzold, S. Mährlein, V. Zbarsky, F. Freimuth, Y. Mokrousov, S. Blügel *et al.*, Terahertz spin current pulses controlled by magnetic heterostructures. Nat. Nanotechnol. **8**, 256 (2013).

[8] S. Sangiao, J. M. De Teresa, L. Morellon, I. Lucas, M. C. Martinez-Velarte, and M. Viret, Control of the spin to charge conversion using the inverse Rashba-Edelstein effect. Appl. Phys. Lett. **106**, 172403 (2015).

[9] W. Zhang, M. B. Jungfleisch, W. Jiang, J. E. Pearson, and A. Hoffmann, Spin pumping and inverse Rashba-Edelstein effect in NiFe/Ag/Bi and NiFe/Ag/Sb. J. Appl. Phys. **117**, 17C727 (2015).

[10] M. Isasa, M. C. Martínez-Velarte, E. Villamor, C. Magén, L. Morellón, J. M. De Teresa, M. R. Ibarra, G. Vignale, E. V. Chulkov, E. E. Krasovskii *et al.*, Origin of inverse Rashba-Edelstein effect detected at the Cu/Bi interface using lateral spin valves. Phys. Rev. B **93**, 014420 (2016).

[11] C. Zhou, Y. P. Liu, Z. Wang, S. J. Ma, M. W. Jia, R. Q. Wu, L. Zhou, W. Zhang, M. K. Liu, Y. Z. Wu *et al.*, Broadband Terahertz Generation via the Interface Inverse Rashba-Edelstein Effect. Phys. Rev. Lett. **121**, 086801 (2018).

[12] T. Seifert, S. Jaiswal, U. Martens, J. Hannegan, L. Braun, P. Maldonado, F. Freimuth, A. Kronenberg, J. Henrizi, I. Radu *et al.*, Efficient metallic spintronic emitters of ultrabroadband terahertz radiation. Nat. Photonics **10**, 483 (2016).

[13] P. Wang, Z. Feng, Y. Yang, D. Zhang, Q. Liu, Z. Xu, Z. Jia, Y. Wu, G. Yu, X. Xu *et al.*, Inverse orbital Hall effect and orbitronic terahertz emission observed in the materials with weak spin-orbit coupling. npj Quantum Mater. **8**, 28 (2023).

[14] S. Kumar ,and S. Kumar, Ultrafast THz probing of nonlocal orbital current in transverse multilayer metallic heterostructures. Nat. Commun. **14**, 8185 (2023).

[15] Y. Xu, F. Zhang, A. Fert, H.-Y. Jaffres, Y. Liu, R. Xu, Y. Jiang, H. Cheng, and W. Zhao, Orbitronics: light-induced orbital currents in Ni studied by terahertz emission experiments. Nat. Commun. **15**, 2043 (2024).

[16] L. Huang, D. Tian, L. Liao, H. Qiu, H. Bai, Q. Wang, F. Pan, C. Zhang, B. Jin, and C. Song, Orbital Current Pumping From Ultrafast Light-driven Antiferromagnetic Insulator. Adv. Mater. **37**, 2402063 (2024).

[17] C. Guo, S. Yin, J. Yang, Y. Ren, P. Ji, Y. Zhang, H. Yan, J. Zhang, B. Cui, Y. Zuo *et al.*, Ferromagnetic Ni enhances terahertz emission through the inverse orbital Hall effect in Ni/Ti bilayers. Phys. Rev. Appl. **24**, 024009 (2025).

[18] L. Liu, T. Jiang, X. Zhao, K. Chen, T. Lai, W. Liu, and Z. Zhang, Qualitative Identification of the Spin-to-Orbital Conversion Mechanism Modulated by Rare-Earth Nd, Gd, and Ho Metals via Terahertz Emission Measurements. Adv. Funct. Mater. **34**, 2411262 (2024).

[19] R. Xu, X. Ning, H. Cheng, Y. Yao, Z. Ren, S. Liu, M. Dai, Y. Xu, S. Li, A. Du *et al.*, Terahertz generation via the inverse orbital Rashba-Edelstein effect at the Ni/$CuO_x$ interface. Phys. Rev. Res. **7**, L012042 (2025).

[20] Y. Liu, Y. Xu, A. Fert, H. Y. Jaffres, T. Nie, S. Eimer, X. Zhang, and W. Zhao, Efficient Orbitronic Terahertz Emission Based on CoPt Alloy. Adv. Mater. **36**, 2404174 (2024).

[21] Y. Zou, Y. Song, Z. Li, J. Zhang, H. Dai, X. Ma, Q. Jin, and Z. Zhang, Enhanced Terahertz Emission via Synergistic Spin-Orbit Conversion in CoPt/Pt/W Multilayers. Adv. Funct. Mater., e15949 (2025).

[22] X. Qiu, W.-S. Noh, K. Narayanapillai, J.-H. Park, K.-J. Lee, H.-W. Lee, H. Yang, Y. Wu, P. Deorani, and D.-H. Yang, Spin-orbit-torque engineering via oxygen manipulation. Nat. Nanotechnol. **10**, 333 (2015).

[23] J. Feng, E. Grimaldi, C. O. Avci, M. Baumgartner, G. Cossu, A. Rossi, and P. Gambardella, Effects of Oxidation of Top and Bottom Interfaces on the Electric, Magnetic, and Spin-Orbit Torque Properties of Pt/Co/$AlO_x$ Trilayers. Phys. Rev. Appl. **13**, 044029 (2020).

[24] K. Hasegawa, Y. Hibino, M. Suzuki, T. Koyama, and D. Chiba, Enhancement of spin-orbit torque by inserting $CoO_x$ layer into Co/Pt interface. Phys. Rev. B **98**, 020405 (2018).

[25] S. Ding, A. Ross, D. Go, L. Baldrati, Z. Ren, F. Freimuth, S. Becker, F. Kammerbauer, J. Yang, G. Jakob *et al.*, Harnessing Orbital-to-Spin Conversion of Interfacial Orbital Currents for Efficient Spin-Orbit Torques. Phys. Rev. Lett. **125**, 177201 (2020).

[26] S. Wu, T. L. Jin, F. N. Tan, C. C. I. Ang, H. Y. Poh, G. J. Lim, and W. S. Lew, Enhancement of spin–orbit torque in Pt/Co/$HfO_x$ heterostructures with voltage-controlled oxygen ion migration. Appl. Phys. Lett. **122**, 122403 (2023).
[27] S. Ding, M.-G. Kang, W. Legrand, and P. Gambardella, Orbital Torque in Rare-Earth Transition-Metal Ferrimagnets. Phys. Rev. Lett. **132**, 236702 (2024).
[28] Z. Zheng, T. Zeng, T. Zhao, S. Shi, L. Ren, T. Zhang, L. Jia, Y. Gu, R. Xiao, H. Zhou *et al.*, Effective electrical manipulation of a topological antiferromagnet by orbital torques. Nat. Commun. **15**, 745 (2024).
[29] S. Ding, P. Noël, G. K. Krishnaswamy, N. Davitti, G. Sala, M. Fantauzzi, A. Rossi, and P. Gambardella, Generation, transmission, and conversion of orbital torque by an antiferromagnetic insulator. Nat. Commun. **16**, 9239 (2025).
[30] See Supplemental Material at [URL] for Sample Preparation and Experimental Setup, Pump fluence dependent THz emission in W/Co/$SiO_2$, THz emission for W($t_W$)/Co(5)/$SiO_2$, Numerical analysis of orbital transport dynamics in W layer, Comparison of THz emission for W/Co/$SiO_2$ and Co/W/$SiO_2$, EELS results for Co(5)/W(5)/$SiO_2$, XPS results for W(5)/Cu(1)/Co(1)/$SiO_2$, Terahertz transmitter performance in W(5)/Co(1)/$SiO_2$, which include Refs. [42-48].
[31] Y. Wu, M. Elyasi, X. Qiu, M. Chen, Y. Liu, L. Ke, and H. Yang, High-Performance THz Emitters Based on Ferromagnetic / Nonmagnetic Heterostructures. Adv. Mater. **29**, 1603031 (2016).
[32] L. Huang, J.-W. Kim, S.-H. Lee, S.-D. Kim, V. M. Tien, K. P. Shinde, J.-H. Shim, Y. Shin, H. J. Shin, S. Kim *et al.*, Direct observation of terahertz emission from ultrafast spin dynamics in thick ferromagnetic films. Appl. Phys. Lett. **115**, 142404 (2019).
[33] T. J. Huisman, R. V. Mikhaylovskiy, J. D. Costa, F. Freimuth, E. Paz, J. Ventura, P. P. Freitas, S. Blügel, Y. Mokrousov, T. Rasing *et al.*, Femtosecond control of electric currents in metallic ferromagnetic heterostructures. Nat. Nanotechnol. **11**, 455 (2016).
[34] T. S. Seifert, S. Jaiswal, J. Barker, S. T. Weber, I. Razdolski, J. Cramer, O. Gueckstock, S. F. Maehrlein, L. Nadvornik, S. Watanabe *et al.*, Femtosecond formation dynamics of the spin Seebeck effect revealed by terahertz spectroscopy. Nat. Commun. **9**, 2899 (2018).

[35] T. Tanaka, H. Kontani, M. Naito, T. Naito, D. S. Hirashima, K. Yamada, and J. Inoue, Intrinsic spin Hall effect and orbital Hall effect in 4*d* and 5*d* transition metals. Phys. Rev. B **77**, 165117 (2008).
[36] H. Kontani, T. Tanaka, D. S. Hirashima, K. Yamada, and J. Inoue, Giant Orbital Hall Effect in Transition Metals: Origin of Large Spin and Anomalous Hall Effects. Phys. Rev. Lett. **102**, 016601 (2009).
[37] M. C. Biesinger, B. P. Payne, A. P. Grosvenor, L. W. M. Lau, A. R. Gerson, and R. S. C. Smart, Resolving surface chemical states in XPS analysis of first row transition metals, oxides and hydroxides: Cr, Mn, Fe, Co and Ni. Appl. Surf. Sci. **257**, 2717 (2011).
[38] S. Haku, A. Musha, T. Gao, and K. Ando, Role of interfacial oxidation in the generation of spin-orbit torques. Phys. Rev. B **102**, 024405 (2020).
[39] F. H. A. Katrib, P. Wehrer, L. Hilaire and G. Maire, The multi-surface structure and catalytic properties of partially reduced $WO_3$, $WO_2$ and $WC+O_2$ or $W+O_2$ as characterized by XPS. J. Electron. Spectrosc. Relat. Phenom. **76**, 195 (1995).
[40] A. P. Shpak, A. M. Korduban, M. M. Medvedskij, and V. O. Kandyba, XPS studies of active elements surface of gas sensors based on $WO_{3-x}$ nanoparticles. J. Electron. Spectrosc. Relat. Phenom. **156**, 172 (2007).
[41] F. Y. Xie, L. Gong, X. Liu, Y. T. Tao, W. H. Zhang, S. H. Chen, H. Meng, and J. Chen, XPS studies on surface reduction of tungsten oxide nanowire film by $Ar^+$ bombardment. J. Electron. Spectrosc. Relat. Phenom. **185**, 112 (2012).
[42] X. Zeng, Q. Wen, Y. Niu, P. Yan, Z. Zhong, and L. Jin, Ballistic-to-diffusive transition of ultrafast orbital current transport in nickel. Phys. Rev. B **112**, 224405 (2025).
[43] S. S. Mishra, J. Lourembam, D. J. X. Lin, and R. Singh, Active ballistic orbital transport in Ni/Pt heterostructure. Nat. Commun. **15**, 4568 (2024).
[44] K. Sriram, R. Mondal, J. Pradhan, A. Haldar, and C. Murapaka, Structural Phase Engineering of (α+β)-W for a Large Spin Hall Angle and Spin Diffusion Length. The Journal of Physical Chemistry C **127**, 22704 (2023).
[45] T. S. Seifert, D. Go, H. Hayashi, R. Rouzegar, F. Freimuth, K. Ando, Y. Mokrousov, and T. Kampfrath, Time-domain observation of ballistic orbital-angular-momentum

currents with giant relaxation length in tungsten. Nat. Nanotechnol. **18**, 1132 (2023).

[46] H. Hayashi, D. Jo, D. Go, T. Gao, S. Haku, Y. Mokrousov, H.-W. Lee, and K. Ando, Observation of long-range orbital transport and giant orbital torque. Commun. Phys. **6**, 32 (2023).

[47] Y. J. Kim, S.-G. Kang, Y. Oh, G. W. Kim, I. H. Cha, H. N. Han, and Y. K. Kim, Microstructural evolution and electrical resistivity of nanocrystalline W thin films grown by sputtering. Mater. Charact. **145**, 473 (2018).

[48] Q. Hao, W. Chen, and G. Xiao, Beta (β) tungsten thin films: Structure, electron transport, and giant spin Hall effect. Appl. Phys. Lett. **106**, 182403 (2015).